\documentclass[aps,prl,twocolumn,amsmath,amssymb,superscriptaddress]{revtex4-2}

\usepackage{graphicx}
\usepackage{bm}
\usepackage{physics}
\usepackage{siunitx}

\usepackage[usenames,dvipsnames]{xcolor} 
\usepackage[
  colorlinks=true,
  citecolor=Blue,
  linkcolor=Blue,
  urlcolor=Blue,
  filecolor=Blue
]{hyperref}
\usepackage[capitalise]{cleveref}

\begin{document}

\title{Persistent coherent quantum dynamics in 2D long-range magnets via magnon binding}

\author{Vighnesh Dattatraya Naik}
\affiliation{%
 Theoretical Physics III, Center for Electronic Correlations and Magnetism,
Institute of Physics, University of Augsburg, 86135 Augsburg, Germany}%

\author{Markus Heyl}
\affiliation{%
 Theoretical Physics III, Center for Electronic Correlations and Magnetism,
Institute of Physics, University of Augsburg, 86135 Augsburg, Germany}%
\affiliation{Centre for Advanced Analytics and Predictive Sciences (CAAPS), University of Augsburg, Universitätsstr. 12a, 86159 Augsburg, Germany}

\date{\today}

\begin{abstract}

The dynamics of 2D long-range quantum magnets represents a current frontier in experimental physics such as in Rydberg atomic systems or trapped ions. In this work we address theoretical challenges in understanding these dynamics by combining large-scale neural quantum state simulations with an effective theory. Our findings uncover a mechanism for persistent coherent quantum dynamics and slow relaxation in 2D long-range quantum magnets. Demonstrated on the 2D transverse-field quantum Ising model with power-law decaying interactions, we observe long-lived oscillatory behavior after quenching the system from a ferromagnetic product state. We explain this phenomenon by the formation of magnon bound states, generated by effective attractive long-range magnon interactions. Our results highlight a generic mechanism for long-lived quantum coherence in 2D quantum magnets that can be directly observed in current quantum simulation platforms.

\end{abstract}

\maketitle
\textit{Introduction.}—
\begin{figure}[t]
  \centering
\includegraphics[width=0.48\textwidth]{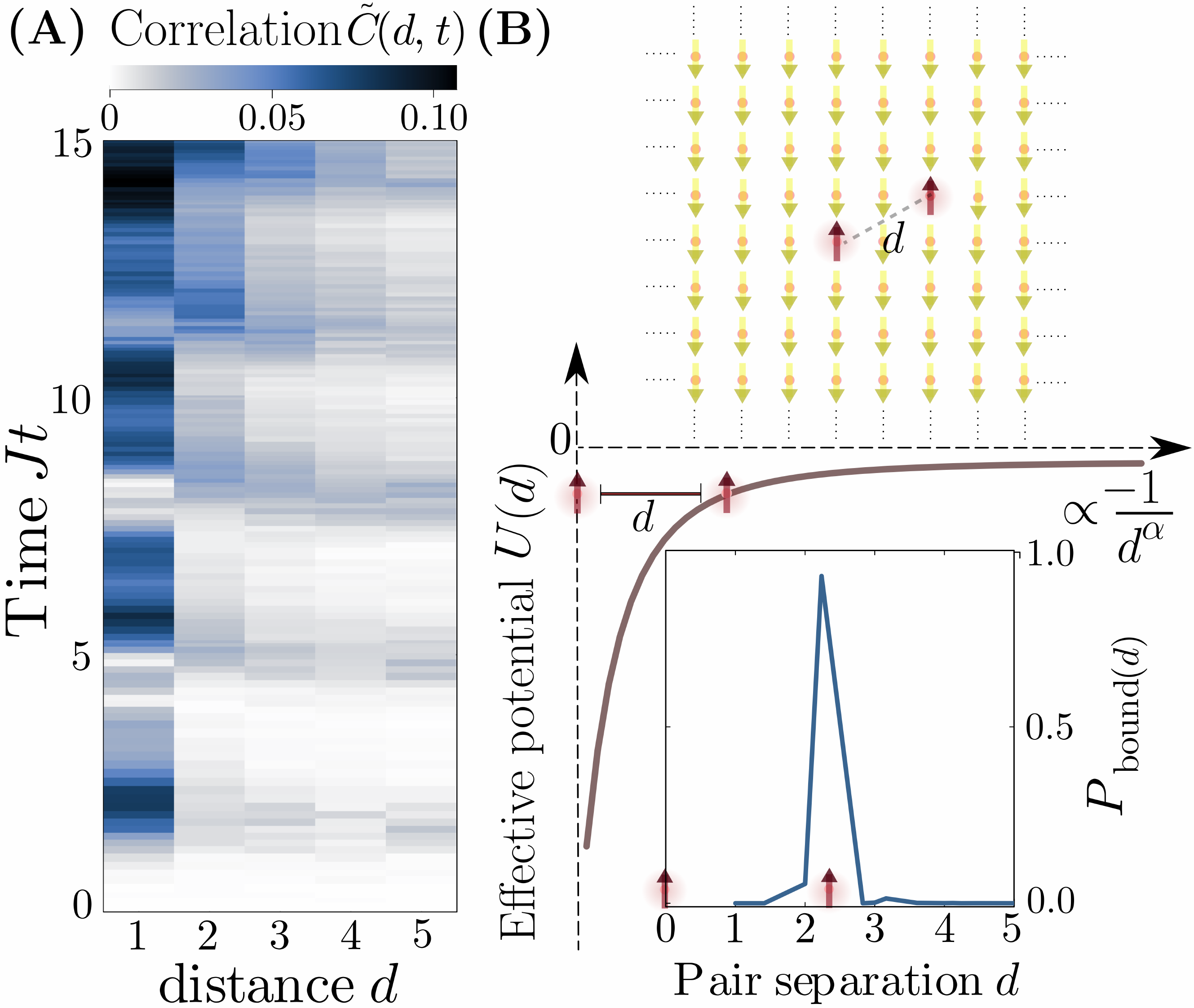}
     \caption{Slow dynamics and magnon binding. (A) Spatiotemporal normalized connected correlator $\tilde{C}(d,t)$ after a quench from $|\!\downarrow\cdots\downarrow\rangle$ on an $11\times11$ lattice ($\alpha=3$, $g/J=0.2$), showing slow dynamics and oscillations. (B) Schematic illustration of two magnons at separation $d$. The lower panel sketches the effective attractive potential $U(d)$ for $\alpha=3$ from Eq.~\ref{eq:U_d}. Inset: the corresponding two-magnon bound-state probability peaked at $\tilde r=\sqrt{5}$ on an $101\times101$ lattice.
    }
  \label{fig:fig1}
\end{figure}
\begin{figure*}[t]
  \centering
  \includegraphics[width=\textwidth]{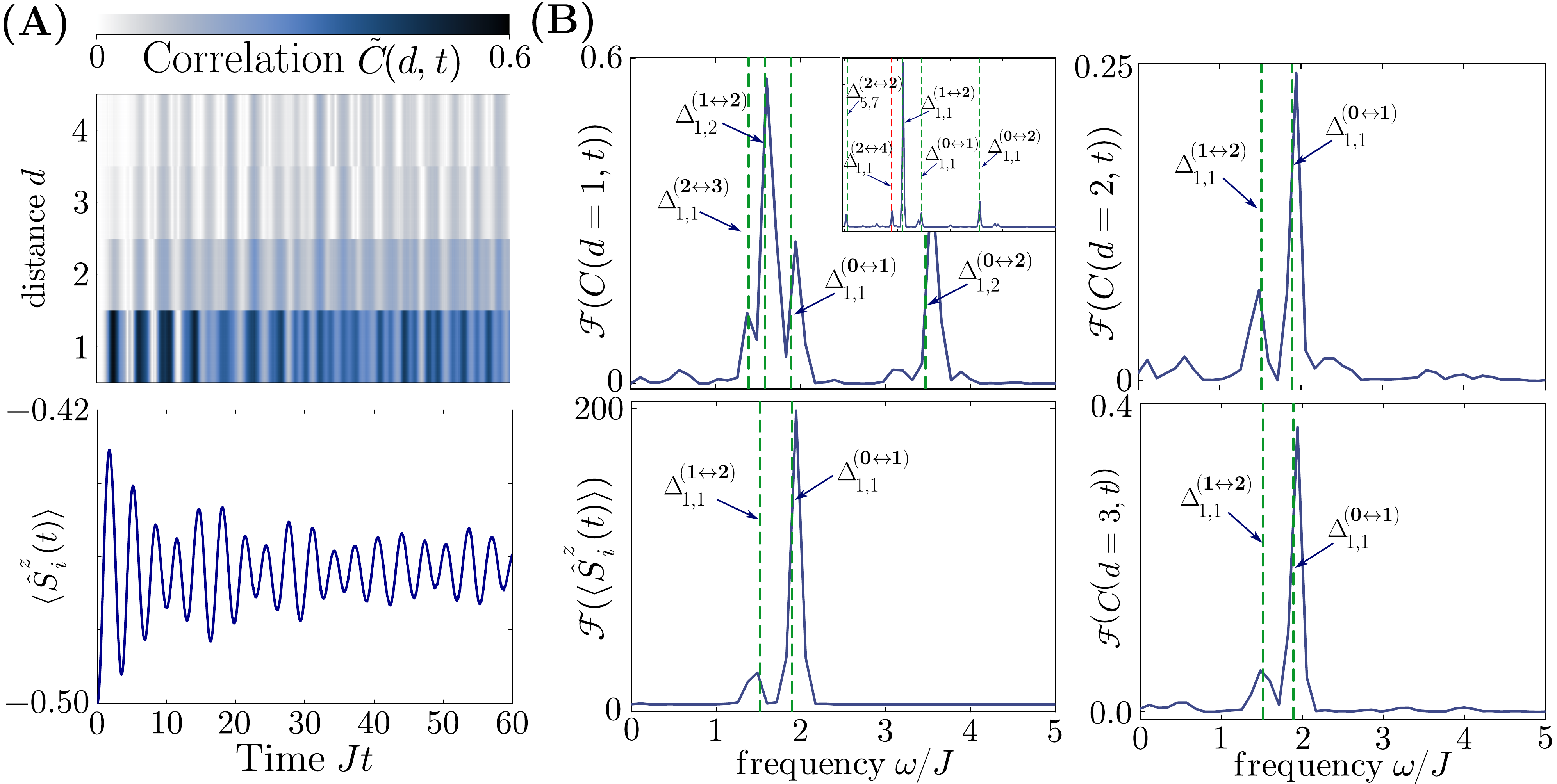}
\caption{Fourier spectroscopy for the post-quench dynamics ($9\times9$, $\alpha = 3$ and $g/J = 0.5$).
(A) Spatiotemporal connected correlator $\tilde{C}(d,t)$, together with the site-averaged longitudinal magnetization $\langle \hat{S}_i^z(t)\rangle$. 
(B) FFTs $\mathcal{F}[\,C(d,t)\,]$ for $d=1,2,3$, and $\mathcal{F}[\,\langle\hat{S}_i^z(t)\rangle\,]$, plotted versus frequency $\omega/J$.
Vertical dashed lines mark energy gaps $\Delta^{(\nu\leftrightarrow \nu')}_{i,j}=E^\nu_i-E^{\nu'}_j$ from effective Hamiltonians restricted to the $\nu$ magnon sectors.
The matching between spectral peaks and energy gaps quantitatively validates the few-body effective description. Inset: Fourier spectra of exact dynamics for a $5\times 5$ system (evolved up to $Jt = 200$).
}
\label{fig:fig2}
\end{figure*}
Long-range quantum magnets arise across atomic, molecular, optical, and solid-state platforms~\cite{DefenuRMP2023}, with prototypical realizations provided by spin models with power-law couplings in trapped ions~\cite{BrittonNature2012,RichermeNature2014,JurcevicNature2014,BohnetScience2016,ZhangNature2017,MonroeRMP2021,Kim2024}, Rydberg atom arrays~\cite{BrowaeysNature2020,EbadiNature2021,BornetNature2023}, and dipolar quantum gases~\cite{YanNature2013,HazzardPRL2014,ChristakisNature2023,PazPRL2013,AlaouiPRL2022}.
In 1D, theoretical and experimental studies have established that long-range interactions modify correlation spreading and the effective light-cone structure~\cite{Hauke2013PRL,RichermeNature2014}, affect entanglement growth~\cite{JurcevicNature2014}, and generate prethermal plateaus and long-lived oscillations that delay thermalization~\cite{LiuPRL2019}.
These phenomena have been linked to emergent quasiparticles, such as domain walls or magnons, their mutual interactions, and the bound states they form~\cite{Bethe1931,Wortis1963,JurcevicNature2014,Kim2024,Kranzl2023,Fukuhara2013,Macri2021PRB}.

However, extending these insights to two dimensions remains a central open problem.
The exponential growth of the Hilbert space together with the rapid buildup of entanglement makes well-established methods such as exact diagonalization and standard tensor-network methods (e.g., tDMRG/MPS) challenging.
Consequently, the quantum dynamics of 2D long-range magnets remain largely unexplored.
Beyond the numerical challenge, the interplay of 2D geometry and power-law interactions can give rise to dynamical mechanisms that are distinct from both the one-dimensional case and the nearest-neighbor limit, raising basic questions:
What are the relevant elementary excitations?
Can nonlocal interactions in 2D bind excitations over several lattice sites?
How does this impact thermalization?

In this Letter, we address these questions by combining large-scale neural quantum state (NQS) simulations~\cite{CarleoScience2017,SchmittPRL2020,SchmittScience2022,schmitt2025} with an effective theoretical description.
We uncover a generic mechanism for persistent coherent quantum dynamics and slow relaxation in 2D long-range quantum magnets.
As an experimentally relevant benchmark, we focus on the 2D transverse-field Ising model with power-law interactions, accessing system sizes and evolution times beyond established numerical approaches.
Quenching from a fully polarized ferromagnetic product state, we observe slow relaxation characterized by persistent, long-lived oscillations, a striking departure from the general expectation of rapid thermalization. 
To interpret this behavior, we construct an effective theory whose eigenspectrum shows excellent agreement with the Fourier spectral analysis of the real-time dynamics.
Crucially, the effective model reveals that long-range interactions mediate a power-law attraction between magnons, yielding multi-magnon bound states even at separations of several lattice sites and producing a large manifold of bound states.
We thus identify extended magnon bound states as the quasiparticles governing slow relaxation in the 2D long-range Ising model, and argue that magnons form bound states whenever long-range attractive interactions compete with weak kinetic terms, making magnon binding a generic mechanism for persistent coherent dynamics in 2D long-range magnets.

\textit{Model.}—
We consider the 2D long-range quantum Ising model described by the Hamiltonian
\begin{equation}
    \hat{H}
    = - \frac{J}{\mathcal{N}_\alpha} 
      \sum_{i\neq j} \frac{\hat{S}_i^z \hat{S}_j^z}{r_{ij}^{\alpha}}
      - g \sum_i \hat{S}_i^x ,
    \label{eq:Hamiltonian}
\end{equation}
where $\hat{S}_i^{\mu}$ ($\mu = x,y,z$) are spin-$1/2$ operators on a square lattice of size $L \times L$ with periodic boundary conditions, and $r_{ij}$ denotes the distance between sites $i$ and $j$.
%
The exponent $\alpha$ controls the range of the interaction: small values of $\alpha$ correspond to slowly decaying couplings, while $\alpha \to \infty$ recovers the nearest-neighbor limit.
We employ Kac normalization
$\mathcal{N}_\alpha = (L^2-1)^{-1}\sum_{{i<j}} r_{ij}^{-\alpha}$, which keeps the energy per site finite as $L \rightarrow \infty$.

\begin{figure*}[t]
  \centering
  \includegraphics[width=\textwidth]{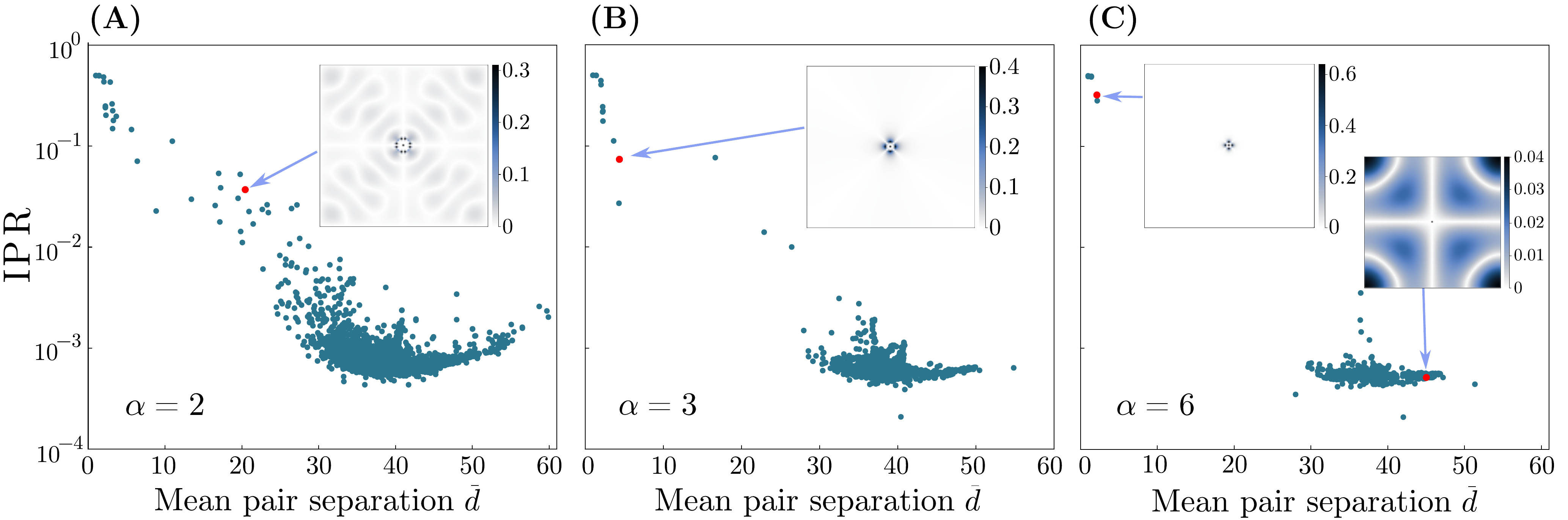}
  \caption{Eigenstate structure in the two-magnon sector ($101 \times 101$, $g/J=0.2$).
  Inverse participation ratio (IPR) versus mean pair separation $\bar d$ for all eigenstates.
  (A) $\alpha=2$: a continuous tail to large $\bar d$ indicates bound states and quasilocalized states.
  (B) $\alpha=3$: intermediate regime with bound states at intermediate separations.
  (C) $\alpha=6$: sharp drop in IPR with $\bar d$ signals short-range binding only.
  Insets: real-space distributions of representative eigenstates in the $(\mathbf{r}_1-\mathbf{r}_2)$ plane.}
  \label{fig:fig3}
\end{figure*}
\textit{Quench protocol and NQS simulations.}—We study the real-time dynamics following a global quench from the fully polarized initial state
$\ket{\psi_0} = \bigotimes_i \ket{\downarrow_i}$ to a finite $g$.
Motivated by the broad experimental availability of dipolar interactions, we consider the dipolar case with exponent $\alpha=3$ as a representative of long-range interactions.
Time evolution is governed by
$\ket{\psi(t)} = e^{-i \hat{H} t} \ket{\psi_0}$, and we monitor the longitudinal magnetization $\langle \hat{S}_i^z(t) \rangle$ and the spin-spin
correlation function
\begin{equation}
    C(d,t) = \langle \hat{S}^z_{i}(t)\hat{S}^z_{i+d}(t) \rangle 
         - \langle \hat{S}^z_{i}(t) \rangle \langle \hat{S}^z_{i+d}(t) \rangle,
\end{equation}
measured along the x-direction of the lattice.
For comparison across different $\alpha$ and $g$, we plot the rescaled correlator $\tilde{C}(d,t)=C(d,t)\cdot\frac{8\mathcal{N}_\alpha J^2}{g^2}$, which provides a suitable normalization across multiple parameter regimes.

To reach system sizes beyond the limits of exact diagonalization for real-time simulations, we employ neural quantum states (NQS), a variational representation of quantum wavefunctions using artificial neural networks.
Within NQS, the time-dependent state $|\psi_{\boldsymbol{\theta(t)}}\rangle=\sum_S \psi_{\boldsymbol{\theta(t)}}(S)|S\rangle$, with $|S\rangle$ denoting the full set of $2^N$ spin configurations, is represented through wave-function amplitudes  $\psi_{\boldsymbol{\theta}(t)}(S)$ which are parametrized by network weights $\boldsymbol{\theta}$.
The evolution of these parameters, $\boldsymbol{\theta}(t)$, is governed by the time-dependent variational principle.
Importantly, NQS provides a numerically exact solution of the dynamics in the sense that convergence is guaranteed upon increasing the size of the underlying artificial neural network.
Specifically, we use a convolutional neural network (CNN) architecture based on the ResNet framework Ref.~\cite{Kaiming2015, chen2024, chen2025}, which is optimized for translational invariance and to capture 2D spatial correlations.
%

Our numerical results, obtained via NQS simulations, show that for an $11\times 11$ lattice [Fig.~\ref{fig:fig1}\textbf{(A)}], the spatiotemporal correlations $\tilde{C}(d,t)$ build up only slowly, exhibiting persistent oscillations at the same time.
Extending the dynamics to times up to $Jt=60$ for a $9\times 9$ lattice [Fig.~\ref{fig:fig2}\textbf{(A)}] reveals that the oscillations in both the longitudinal magnetization $\langle \hat{S}_i^z(t) \rangle$ and the correlation $\tilde{C}(d,t)$ are unconventionally long-lived.
To resolve their characteristic frequencies, we perform a discrete Fourier transform $\mathcal{F}$ of $C(d,t)$ and $\langle \hat{S}_i^z(t)\rangle$, applying a Hamming window over the interval $Jt=5$–$60$.
%
The resulting spectra exhibit sharp peaks [Fig.~\ref{fig:fig2}\textbf{(B)}], signaling the emergence of well-defined modes.
To identify their microscopic origin and account for the slow dynamics, we next develop an effective low-energy theory.

\textit{Effective few-body theory.}—We begin by identifying the low-energy excitations of the unperturbed Hamiltonian $H_0=- \frac{J}{\mathcal{N}_\alpha}\sum_{i\neq j} \frac{\hat{S}_i^z \hat{S}_j^z}{r_{ij}^{\alpha}}$, with fully polarized initial state $\ket{\downarrow\downarrow\cdots\downarrow}$ as one of the two lowest energy states.
Excitations above this background can be described in terms of magnons (or spin flips).
For a configuration containing $\nu$ magnons at lattice sites $\{i_1, \cdots,i_\nu\}$, the energy relative to the ground state energy $E_0$ is
\begin{equation*}
    E_{\nu}(\{i_m\}) - E_{0} = 2\nu J\left(1 - \frac{1}{L^2}\right) - \frac{2J}{\mathcal{N}_\alpha}\sum_{1\le k < l\le\nu} \frac{1}{r^\alpha_{i_ki_l}}.
\end{equation*}
The first term, which grows linearly with the number of magnons, represents the cost of an isolated magnon. 
The second term reflects an attractive interaction energy between all magnon pairs.

In the nearest-neighbor limit ($\alpha=\infty$), the interaction term vanishes unless $r_{i_ki_l}=1$.
The energy expression then reduces to a constant proportional to the number of domain-wall links, implying that the excitation spectrum is classified solely by the interface length of spin-flip domains.
For finite $\alpha$, however, the long-range couplings generate a much richer structure.
Specifically, the single-magnon sector forms the lowest excited band, followed by the two-magnon manifold.
Higher sectors contain states with three or more magnons, but their ordering is highly sensitive to the geometry of how the magnons are arranged.
For example, a compact four-magnon cluster lies energetically below a configuration of three well-separated magnons.
Within this hierarchy, the states directly above the two-magnon manifold are those three-magnon configurations in which at least two magnons occupy nearest-neighbor sites.

Having established the structure of low-energy excitations in the pure Ising Hamiltonian, we now focus on the influence of the perturbation $V=-g\sum_{i}\hat{S}^x_i$.
Since, for finite $\alpha$, the $\nu=0,1,2$ sectors together with the three-magnon sector containing at least a nearest-neighbor pair constitute well-separated excitation bands with gaps of $\mathcal{O}(J)$, the effect of a weak transverse field strength $g$ can be treated perturbatively.
Applying a Schrieffer–Wolff (SWT) transformation up to second order in $g$ in each sector, we integrate out couplings to other sectors at leading order and thereby obtain an effective Hamiltonian acting solely within the sector with corrections to this structure arising only at order $\mathcal{O}(g^3/J^2)$.
For the two-magnon manifold, imposing translational symmetry and restricting to the zero-momentum subspace,
yields the effective Hamiltonian
\begin{multline}
\hat H^{(2)}_{\mathrm{eff}} = E_2 + 
\sum_{\mathbf d}\!U(\mathbf d) \ket{\mathbf d}\!\bra{\mathbf d} + \sum_{\mathbf d \neq \mathbf d'} t_{\mathbf d,\mathbf d'} \ket{\mathbf d'}\!\bra{\mathbf d}.
\end{multline}
where the basis states $|\mathbf d\rangle$ represent two magnons separated by the relative coordinate $\mathbf d$.
$E_2 \approx E_0 + 4J - g^2L^2/8J$ fixes the overall energy of the two-magnon manifold.
A central finding of our effective description is that two magnons experience an attractive interaction potential of long-range character,
\begin{align} 
\ U(\mathbf d) &= - \left(2J - \frac{g^2}{2J}\right) \frac{1}{\mathcal N_\alpha |\mathbf d|^\alpha} + \delta U(\mathbf d),
\label{eq:U_d}
\end{align}
with $\delta U(\mathbf d)$ denoting corrections proportional to $g^2/J$, which at large distances $|\mathbf{d}|$ provide only a subleading contribution to the attractive interactions proportional to $\frac{-g^2}{8J\mathcal{N}_\alpha}\frac{1}{|\mathbf{d}|^{2(\alpha-1)}}$, as long as $\alpha>2$.
The off-diagonal amplitudes $t_{\mathbf d,\mathbf d'}$ are real and symmetric, describing correlated long-range pair hoppings, which at large $|\mathbf{d}-\mathbf{d'}|$ are given by $t_{\mathbf d,\mathbf d'} \approx \frac{-g^2}{8J\mathcal{N}_\alpha} \frac{1}{|\mathbf d - \mathbf d'|^{\alpha}}$ (explicit derivations are provided in the supplementary material).
For all numerical calculations, however, we employ the full (non-asymptotic) effective Hamiltonian derived via the Schrieffer–Wolff transformation.
Notice that while an attractive interaction potential might be naturally associated with the formation of bound states in the case of short-range kinetics~\cite{yang1989, chadan2003}, we find that the interplay between the long-range attractive potential and the correlated pair long-range hoppings produces a rich spectrum of two-magnon bound states.

\begin{figure}[t]
    \centering
    \includegraphics[width=\columnwidth]{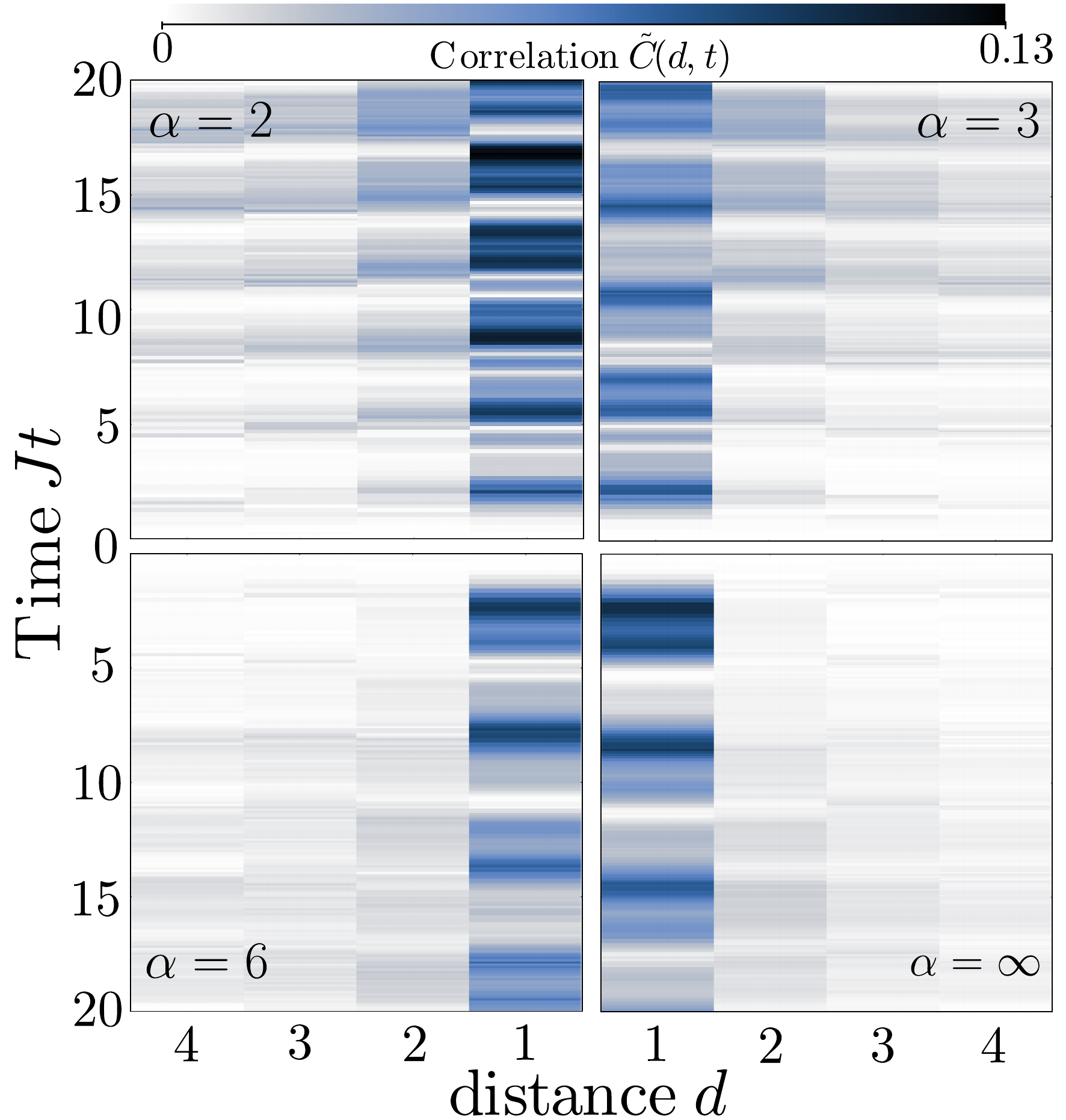}%
    \caption{$\tilde{C}(d,t)$ for a $9\times9$ lattice with $g/J=0.2$. The panels compare interaction exponents
        $\alpha = 2,3$ with $\alpha = 6$ and the nearest-neighbor limit $\alpha = \infty$.
    }
    \label{fig:fig4}
\end{figure}

To validate this theoretical framework, we diagonalize $\hat H_{\mathrm{eff}}^{(\nu)}$, and then compare the resulting level spacings with peak positions extracted from the Fourier spectra of $\langle \hat{S}^z_i(t) \rangle$ and $C(d,t)$.
Figure~\ref{fig:fig2}\textbf{(B)} shows these spectra for $9\times 9$ lattice with $g/J=0.5$ and $\alpha=3$ overlaid with excitation gaps $\Delta^{(\nu\leftrightarrow\nu')}_{i,j} = |E^\nu_i - E_j^{\nu'}|$, where $E^\nu_i$ denotes the $i-th$ energy level of $\hat H_{\mathrm{eff}}^{(\nu)}$.
The dominant peak at $\omega/J \simeq 1.89$ corresponds to $\Delta^{(0\leftrightarrow 1)}_{1,1}$, i.e., creating a single magnon above the polarized background (see Supplementary Material for the single-magnon dispersion).
Additional peaks arise from $\Delta^{(0\leftrightarrow 2)}$ and $\Delta^{(1\leftrightarrow 2)}$, corresponding to creating a two-magnon bound state out of the polarized state or out of the single-magnon sector, respectively.
We further resolve a peak associated with a three-magnon process $\Delta^{(1\leftrightarrow 3)}$, corresponding to the creation of three tightly bound magnons.
Some peaks appear slightly offset from the exact gap values due to the limited frequency resolution imposed by the NQS time window. 
As a result, the discrete frequency grid does not include points exactly at those gaps, so the spectral weight appears at the nearest available frequency bin, leading to a small apparent mismatch.

To achieve a more refined frequency resolution, the inset displays the FFT of the exact dynamics for a \(5\times 5\) system evolved up to $Jt = 200$. 
The resulting spectrum agrees very well with the effective predictions, except for the red line. 
This deviation arises because, for $L\le5$, the excitations above the two-magnon manifold, those involving three or more magnons, do not form a well-separated band, which limits the applicability of the SWT description. 
Nevertheless, exact diagonalization reveals that this peak originates from a compact four-magnon cluster. 
These observations show that the slow dynamics are controlled by a sparse set of well-defined few-magnon excitations rather than a broad many-body continuum. 
Furthermore, the clear one-to-one correspondence between the spectral peaks extracted from the NQS dynamics and the energy gaps of the effective theory confirms the accuracy of both approaches up to at least $g/J=0.5$.

We emphasize that the effective Hamiltonian is broadly applicable to 2D long-range magnets.
Whenever a microscopic model features power-law decaying ferromagnetic couplings, the low-energy sector above a fully polarized ferromagnetic state is governed by few-magnon excitations.
Adding kinetic terms such as spin flips, hopping, or pair creation and annihilation, or any combination thereof, yields, under a SW transformation, an effective Hamiltonian of very similar structure.

\textit{Emergence of magnon bound states.}—Having validated the effective theory with our exact numerical simulations with NQS, we now aim to analyze in more detail the two-magnon eigenstates.
In this section, we address two central questions: 
(i)~whether the long-range attractive potential binds two magnons beyond the nearest-neighbor separation,
(ii)~how the spatial structure of these states depends on the interaction exponent~$\alpha$.
To characterize these states, we analyze the eigenstates $|\psi(\mathbf d)\rangle$ of the effective two-magnon Hamiltonian $\hat{H}_{\text{eff}}^{(2)}$.
For each eigenstate, we compute its inverse participation ratio (IPR), \(\mathrm{IPR}[\psi]=\sum_{\mathbf d}|\psi(\mathbf d)|^{4}\), which distinguishes localized bound pairs (high IPR) from delocalized scattering states (low IPR). 
Intermediate IPR values correspond to quasilocalized states, whose wavefunctions contain a localized core accompanied by a weak delocalized tail.
We also compute the mean pair separation, \(\bar d=\sum_{\mathbf d} |\psi(\mathbf d)|^{2}\,|\mathbf d|\), to determine the average relative positions.
Figure \ref{fig:fig3} shows the resulting IPR-$\bar d$ distribution for several power-law exponents $\alpha$ on a $101 \times 101$ lattice with $g/J=0.2$, together with representative real-space probability densities in the relative coordinate $(\mathbf{r}_1 - \mathbf{r}_2)$.

For strongly long-ranged interactions ($\alpha=2$), many eigenstates exhibit a large IPR for mean separations up to $\bar{d}=4$, followed by a continuous decay of IPR with increasing $\bar{d}$ [Fig.~\ref{fig:fig3}\textbf{(A)}].
This indicates that the spectrum contains bound magnon pairs with separations as large as $d=4$.
Furthermore, a substantial fraction of eigenstates displays quasilocalized character, with intermediate values of both IPR and $\bar{d}$, with a localized component that persists out to separations of $d=6$.
For $\alpha=3$, the number of quasilocalized states is noticeably reduced [Fig.~\ref{fig:fig3}\textbf{(B)}], although bound pairs still survive up to $\bar d \approx 4$. 
These findings demonstrate that long-range interactions stabilize magnons at distances well beyond nearest neighbors and also enhance the number of localized and quasilocalized states.
By contrast, in the short-range regime ($\alpha=6$), only a small number of eigenstates exhibit high IPR, while the majority show low IPR at large $\bar{d}$.
Consequently, the magnon binding is restricted to short separations up to $d \approx \sqrt 2$ [Fig.~\ref{fig:fig3}\textbf{(C)}].

Notice that within the approximation, the effective interaction \(U(\mathbf d)\) in Eq.~\ref{eq:U_d} becomes less attractive as \(g\) increases, which correspondingly suppresses the formation of bound states at larger separations.
For example, at \(g/J=0.5\) we find that for \(\alpha = 2\) and \(\alpha = 3\), localized states persist up to separations \(d = 2\) and \(d = \sqrt{2}\), respectively. For \(\alpha = 6\) and \(\alpha = \infty\),  only nearest-neighbor bound states remain.

Figure~\ref{fig:fig4} shows the time evolution of the correlations $\tilde{C}(d, t)$ for various interaction exponents $\alpha$ at $g/J=0.2$. 
A clear distinction emerges between the short-range regime ($\alpha = 6,\infty$) 
and the long-range regime ($\alpha = 2,3$). 
For large $\alpha$, the signal is stronger at the nearest-neighbor distance $d=1$, and beyond this separation, the dynamics becomes nearly monochromatic, consistent with the presence of only tight bound magnon pairs and a few characteristic oscillation frequencies.
In contrast, for small $\alpha$, the correlator exhibits multiple frequency components and a significant signal at $d>1$, reflecting the existence of several extended bound states.

\textit{Discussion and Outlook.}
Our NQS simulations and effective theory establish magnon binding as a generic microscopic mechanism for persistent coherent dynamics and slow relaxation in 2D long-range magnets.
Future experiments with Rydberg arrays and trapped ions could directly probe these spatially extended bound magnons using site-resolved correlation spectroscopy, enabling real-time visualization of bound-pair formation and decay.
This mechanism is fundamentally distinct from the 1D case, where domain walls are the elementary excitations and long-range interactions confine domain-wall kinks~\cite{LiuPRL2019}, and from the 2D nearest-neighbor limit, where slow dynamics has been linked to approximate Hilbert-space fragmentation due to the near-conservation of domain-wall length in the weak-field regime~\cite{Yoshinaga2022,Balducci2022,Balducci2023,Tindall2024PRL,Luka2025PRB}.
Beyond the transverse-field Ising model studied here, our framework, in principle, applies broadly to 2D long-range magnets with diverse kinetic processes.
Accordingly, we expect that magnon binding could be a generic feature of 2D long-range quantum magnets, potentially also in 3D, where magnons again constitute the elementary excitations.

\textit{Data availability.}—Data to reproduce all figures are available at Zenodo~\cite{dataset_long-range_quantum_Ising_model}.

\textit{Acknowledgements.}—The NQS dynamics is simulated using the 
\texttt{jVMC} package~\cite{Schmitt2022}, and the exact dynamics are performed with the \texttt{QuSpin} package~\cite{Weinberg2017}. This project has received 
funding from the European Research Council (ERC) under the European Union’s 
Horizon 2020 research and innovation programme (Grant Agreement No.~853443). 
This work was supported by the German Research Foundation (DFG) via 
project 492547816 (TRR~360). We gratefully acknowledge the scientific support 
and high-performance computing resources provided by the LiCCA HPC cluster at 
the University of Augsburg and by the Erlangen National High Performance 
Computing Center (NHR) of the Friedrich-Alexander-Universität Erlangen–Nürnberg 
(NHR Project No.~nqsQuMat). The LiCCA cluster is co-funded by the DFG 
(Project ID 499211671). NHR funding is provided by federal and Bavarian state 
authorities, and the NHR@FAU hardware is partially funded by the DFG (Grant 
No.~440719683).

\textit{Note added.}—During the preparation of this manuscript, we became aware 
of related work by Kaltenmark \emph{et al.}~\cite{kaltenmark2025}, who study 
spectroscopic signatures in a kinetically constrained long-range interacting 
two-dimensional spin system. While their microscopic model and parameter regime 
differ from ours, both works highlight the emergence of nontrivial elementary 
excitations in long-range interacting quantum systems.

\nocite{*}
\bibliography{References}

@article{DefenuRMP2023,
  title = {Long-range interacting quantum systems},
  author = {Defenu, Nicol\`o and Donner, Tobias and Macr\`{\i}, Tommaso and Pagano, Guido and Ruffo, Stefano and Trombettoni, Andrea},
  journal = {Rev. Mod. Phys.},
  volume = {95},
  issue = {3},
  pages = {035002},
  numpages = {70},
  year = {2023},
  month = {Aug},
  publisher = {American Physical Society},
  doi = {10.1103/RevModPhys.95.035002},
  url = {https://link.aps.org/doi/10.1103/RevModPhys.95.035002}
}

@article{BrittonNature2012,
  title={Engineered two-dimensional Ising interactions in a trapped-ion quantum simulator with hundreds of spins},
  author={Britton, Joseph W and Sawyer, Brian C and Keith, Adam C and Wang, C-C Joseph and Freericks, James K and Uys, Hermann and Biercuk, Michael J and Bollinger, John J},
  journal={Nature},
  volume={484},
  number={7395},
  pages={489--492},
  year={2012},
  publisher={Nature Publishing Group UK London},
  doi = {https://doi.org/10.1038/nature10981}
}

@article{RichermeNature2014,
  title={Non-local propagation of correlations in quantum systems with long-range interactions},
  author={Richerme, Philip and Gong, Zhe-Xuan and Lee, Aaron and Senko, Crystal and Smith, Jacob and Foss-Feig, Michael and Michalakis, Spyridon and Gorshkov, Alexey V and Monroe, Christopher},
  journal={Nature},
  volume={511},
  number={7508},
  pages={198--201},
  year={2014},
  publisher={Nature Publishing Group UK London},
  doi = {https://doi.org/10.1038/nature13450}
}

@article{JurcevicNature2014,
  title={Quasiparticle engineering and entanglement propagation in a quantum many-body system},
  author={Jurcevic, Petar and Lanyon, Ben P and Hauke, Philipp and Hempel, Cornelius and Zoller, Peter and Blatt, Rainer and Roos, Christian F},
  journal={Nature},
  volume={511},
  number={7508},
  pages={202--205},
  year={2014},
  publisher={Nature Publishing Group UK London},
  doi = {https://doi.org/10.1038/nature13461}
}

@article{BohnetScience2016,
  title={Quantum spin dynamics and entanglement generation with hundreds of trapped ions},
  author={Bohnet, Justin G and Sawyer, Brian C and Britton, Joseph W and Wall, Michael L and Rey, Ana Maria and Foss-Feig, Michael and Bollinger, John J},
  journal={Science},
  volume={352},
  number={6291},
  pages={1297--1301},
  year={2016},
  publisher={American Association for the Advancement of Science},
  doi = {https://doi.org/10.1126/science.aad9958}
}

@article{ZhangNature2017,
  title={Observation of a many-body dynamical phase transition with a 53-qubit quantum simulator},
  author={Zhang, Jiehang and Pagano, Guido and Hess, Paul W and Kyprianidis, Antonis and Becker, Patrick and Kaplan, Harvey and Gorshkov, Alexey V and Gong, Z-X and Monroe, Christopher},
  journal={Nature},
  volume={551},
  number={7682},
  pages={601--604},
  year={2017},
  publisher={Nature Publishing Group UK London},
  doi = {https://doi.org/10.1038/nature24654}
}

@article{MonroeRMP2021,
  title = {Programmable quantum simulations of spin systems with trapped ions},
  author = {Monroe, C. and Campbell, W. C. and Duan, L.-M. and Gong, Z.-X. and Gorshkov, A. V. and Hess, P. W. and Islam, R. and Kim, K. and Linke, N. M. and Pagano, G. and Richerme, P. and Senko, C. and Yao, N. Y.},
  journal = {Rev. Mod. Phys.},
  volume = {93},
  issue = {2},
  pages = {025001},
  numpages = {57},
  year = {2021},
  month = {Apr},
  publisher = {American Physical Society},
  doi = {10.1103/RevModPhys.93.025001},
  url = {https://link.aps.org/doi/10.1103/RevModPhys.93.025001}
}

@article{YanNature2013,
  title={Observation of dipolar spin-exchange interactions with lattice-confined polar molecules},
  author={Yan, Bo and Moses, Steven A and Gadway, Bryce and Covey, Jacob P and Hazzard, Kaden RA and Rey, Ana Maria and Jin, Deborah S and Ye, Jun},
  journal={Nature},
  volume={501},
  number={7468},
  pages={521--525},
  year={2013},
  publisher={Nature Publishing Group UK London},
  doi = {https://doi.org/10.1038/nature12483}
}

@article{HazzardPRL2014,
  title = {Many-Body Dynamics of Dipolar Molecules in an Optical Lattice},
  author = {Hazzard, Kaden R. A. and Gadway, Bryce and Foss-Feig, Michael and Yan, Bo and Moses, Steven A. and Covey, Jacob P. and Yao, Norman Y. and Lukin, Mikhail D. and Ye, Jun and Jin, Deborah S. and Rey, Ana Maria},
  journal = {Phys. Rev. Lett.},
  volume = {113},
  issue = {19},
  pages = {195302},
  numpages = {5},
  year = {2014},
  month = {Nov},
  publisher = {American Physical Society},
  doi = {10.1103/PhysRevLett.113.195302},
  url = {https://link.aps.org/doi/10.1103/PhysRevLett.113.195302}
}

@article{ChristakisNature2023,
  title={Probing site-resolved correlations in a spin system of ultracold molecules},
  author={Christakis, Lysander and Rosenberg, Jason S and Raj, Ravin and Chi, Sungjae and Morningstar, Alan and Huse, David A and Yan, Zoe Z and Bakr, Waseem S},
  journal={Nature},
  volume={614},
  number={7946},
  pages={64--69},
  year={2023},
  publisher={Nature Publishing Group UK London},
  doi = {https://doi.org/10.1038/s41586-022-05558-4}
}

@article{PazPRL2013,
  title = {Nonequilibrium Quantum Magnetism in a Dipolar Lattice Gas},
  author = {de Paz, A. and Sharma, A. and Chotia, A. and Mar\'echal, E. and Huckans, J. H. and Pedri, P. and Santos, L. and Gorceix, O. and Vernac, L. and Laburthe-Tolra, B.},
  journal = {Phys. Rev. Lett.},
  volume = {111},
  issue = {18},
  pages = {185305},
  numpages = {5},
  year = {2013},
  month = {Oct},
  publisher = {American Physical Society},
  doi = {10.1103/PhysRevLett.111.185305},
  url = {https://link.aps.org/doi/10.1103/PhysRevLett.111.185305}
}

@article{AlaouiPRL2022,
  title = {Measuring Correlations from the Collective Spin Fluctuations of a Large Ensemble of Lattice-Trapped Dipolar Spin-3 Atoms},
  author = {Alaoui, Youssef Aziz and Zhu, Bihui and Muleady, Sean Robert and Dubosclard, William and Roscilde, Tommaso and Rey, Ana Maria and Laburthe-Tolra, Bruno and Vernac, Laurent},
  journal = {Phys. Rev. Lett.},
  volume = {129},
  issue = {2},
  pages = {023401},
  numpages = {6},
  year = {2022},
  month = {Jul},
  publisher = {American Physical Society},
  doi = {10.1103/PhysRevLett.129.023401},
  url = {https://link.aps.org/doi/10.1103/PhysRevLett.129.023401}
}

@article{BrowaeysNature2020,
  title={Many-body physics with individually controlled Rydberg atoms},
  author={Browaeys, Antoine and Lahaye, Thierry},
  journal={Nature Physics},
  volume={16},
  number={2},
  pages={132--142},
  year={2020},
  publisher={Nature Publishing Group UK London},
  doi={https://doi.org/10.1038/s41567-019-0733-z}
}

@article{EbadiNature2021,
  title={Quantum phases of matter on a 256-atom programmable quantum simulator},
  author={Ebadi, Sepehr and Wang, Tout T and Levine, Harry and Keesling, Alexander and Semeghini, Giulia and Omran, Ahmed and Bluvstein, Dolev and Samajdar, Rhine and Pichler, Hannes and Ho, Wen Wei and others},
  journal={Nature},
  volume={595},
  number={7866},
  pages={227--232},
  year={2021},
  publisher={Nature Publishing Group UK London},
  doi={https://doi.org/10.1038/s41586-021-03582-4}
}

@article{BornetNature2023,
  title={Scalable spin squeezing in a dipolar Rydberg atom array},
  author={Bornet, Guillaume and Emperauger, Gabriel and Chen, Cheng and Ye, Bingtian and Block, Maxwell and Bintz, Marcus and Boyd, Jamie A and Barredo, Daniel and Comparin, Tommaso and Mezzacapo, Fabio and others},
  journal={Nature},
  volume={621},
  number={7980},
  pages={728--733},
  year={2023},
  publisher={Nature Publishing Group UK London},
  doi = {https://doi.org/10.1038/s41586-023-06414-9}
}

@article{Bethe1931,
  author = {Bethe, H.},
  year = {1931},
  title = {Zur Theorie der Metalle},
  journal = {Zeitschrift für Physik},
  pages = {205},
  ep = {226},
  volume = {71},
  number = {3},
  sn = {0044-3328},
  url = {https://doi.org/10.1007/BF01341708},
  doi = {10.1007/BF01341708},
  id = {Bethe1931},
}

@article{Wortis1963,
  title = {Bound States of Two Spin Waves in the Heisenberg Ferromagnet},
  author = {Wortis, Michael},
  journal = {Phys. Rev.},
  volume = {132},
  issue = {1},
  pages = {85--97},
  numpages = {0},
  year = {1963},
  month = {Oct},
  publisher = {American Physical Society},
  doi = {10.1103/PhysRev.132.85},
  url = {https://link.aps.org/doi/10.1103/PhysRev.132.85}
}

@article{Kranzl2023,
  title = {Observation of Magnon Bound States in the Long-Range, Anisotropic Heisenberg Model},
  author = {Kranzl, Florian and Birnkammer, Stefan and Joshi, Manoj K. and Bastianello, Alvise and Blatt, Rainer and Knap, Michael and Roos, Christian F.},
  journal = {Phys. Rev. X},
  volume = {13},
  issue = {3},
  pages = {031017},
  numpages = {12},
  year = {2023},
  month = {Aug},
  publisher = {American Physical Society},
  doi = {10.1103/PhysRevX.13.031017},
  url = {https://link.aps.org/doi/10.1103/PhysRevX.13.031017}
}

@article{Fukuhara2013,
  author = {Fukuhara, Takeshi and Schauß, Peter and Endres, Manuel and Hild, Sebastian and Cheneau, Marc and Bloch, Immanuel and Gross, Christian},
  year = {2013},
  da = {2013/10/01},
  title = {Microscopic observation of magnon bound states and their dynamics},
  journal = {Nature},
  pages = {76},
  volume = {502},
  number = {7469},
  sn = {1476-4687},
  url = {https://doi.org/10.1038/nature12541},
  doi = {10.1038/nature12541},
  id = {Fukuhara2013},
}

@article{Kim2024,
  title = {Realization of an Extremely Anisotropic Heisenberg Magnet in Rydberg Atom Arrays},
  author = {Kim, Kangheun and Yang, Fan and M\o{}lmer, Klaus and Ahn, Jaewook},
  journal = {Phys. Rev. X},
  volume = {14},
  issue = {1},
  pages = {011025},
  numpages = {13},
  year = {2024},
  month = {Feb},
  publisher = {American Physical Society},
  doi = {10.1103/PhysRevX.14.011025},
  url = {https://link.aps.org/doi/10.1103/PhysRevX.14.011025}
}

@article{Hauke2013PRL,
  author  = {Hauke, Philipp and Tagliacozzo, Luca},
  title   = {Spread of Correlations in Long-Range Interacting Quantum Systems},
  journal = {Phys. Rev. Lett.},
  volume  = {111},
  pages   = {207202},
  year    = {2013},
  doi     = {10.1103/PhysRevLett.111.207202}
}

@article{LiuPRL2019,
  title = {Confined Quasiparticle Dynamics in Long-Range Interacting Quantum Spin Chains},
  author = {Liu, Fangli and Lundgren, Rex and Titum, Paraj and Pagano, Guido and Zhang, Jiehang and Monroe, Christopher and Gorshkov, Alexey V.},
  journal = {Phys. Rev. Lett.},
  volume = {122},
  issue = {15},
  pages = {150601},
  numpages = {7},
  year = {2019},
  month = {Apr},
  publisher = {American Physical Society},
  doi = {10.1103/PhysRevLett.122.150601},
  url = {https://link.aps.org/doi/10.1103/PhysRevLett.122.150601}
}

@article{Macri2021PRB,
  title = {Bound state dynamics in the long-range spin-$\frac{1}{2}$ XXZ model},
  author = {Macr\`{\i}, T. and Lepori, L. and Pagano, G. and Lewenstein, M. and Barbiero, L.},
  journal = {Phys. Rev. B},
  volume = {104},
  issue = {21},
  pages = {214309},
  numpages = {13},
  year = {2021},
  month = {Dec},
  publisher = {American Physical Society},
  doi = {10.1103/PhysRevB.104.214309},
  url = {https://link.aps.org/doi/10.1103/PhysRevB.104.214309}
}

@article{Tindall2024PRL,
  title = {Confinement in the Transverse Field Ising Model on the Heavy Hex Lattice},
  author = {Tindall, Joseph and Sels, Dries},
  journal = {Phys. Rev. Lett.},
  volume = {133},
  issue = {18},
  pages = {180402},
  numpages = {8},
  year = {2024},
  month = {Oct},
  publisher = {American Physical Society},
  doi = {10.1103/PhysRevLett.133.180402},
  url = {https://link.aps.org/doi/10.1103/PhysRevLett.133.180402}
}

@article{Luka2025PRB,
  title = {Constrained dynamics and confinement in the two-dimensional quantum Ising model},
  author = {Pave\ifmmode \check{s}\else \v{s}\fi{}i\ifmmode \acute{c}\else \'{c}\fi{}, Luka and Jaschke, Daniel and Montangero, Simone},
  journal = {Phys. Rev. B},
  volume = {111},
  issue = {14},
  pages = {L140305},
  numpages = {7},
  year = {2025},
  month = {Apr},
  publisher = {American Physical Society},
  doi = {10.1103/PhysRevB.111.L140305},
  url = {https://link.aps.org/doi/10.1103/PhysRevB.111.L140305}
}

@article{CarleoScience2017,
	author = {Giuseppe Carleo and Matthias Troyer},
	doi = {10.1126/science.aag2302},
	journal = {Science},
	number = {6325},
	pages = {602-606},
	title = {Solving the quantum many-body problem with artificial neural networks},
	volume = {355},
	year = {2017},
}

@article{SchmittPRL2020,
  title = {Quantum Many-Body Dynamics in Two Dimensions with Artificial Neural Networks},
  author = {Schmitt, Markus and Heyl, Markus},
  journal = {Phys. Rev. Lett.},
  volume = {125},
  issue = {10},
  pages = {100503},
  numpages = {7},
  year = {2020},
  month = {Sep},
  publisher = {American Physical Society},
  doi = {10.1103/PhysRevLett.125.100503},
  url = {https://link.aps.org/doi/10.1103/PhysRevLett.125.100503}
}

@article{SchmittScience2022,
	author = {Markus Schmitt and Marek M. Rams and Jacek Dziarmaga and Markus Heyl and Wojciech H. Zurek},
	journal = {Science Advances},
	number = {37},
	pages = {eabl6850},
	title = {Quantum phase transition dynamics in the two-dimensional transverse-field Ising model},
    doi={https://doi.org/10.1126/sciadv.abl6850},
	volume = {8},
	year = {2022}}

@misc{chen2025,
      title={Convolutional transformer wave functions}, 
      author={Ao Chen and Vighnesh Dattatraya Naik and Markus Heyl},
      year={2025},
      eprint={2503.10462},
      archivePrefix={arXiv},
      primaryClass={cond-mat.dis-nn},
      url={https://arxiv.org/abs/2503.10462}, 
}

@misc{Kaiming2015,
      title={Deep Residual Learning for Image Recognition}, 
      author={Kaiming He and Xiangyu Zhang and Shaoqing Ren and Jian Sun},
      year={2015},
      eprint={1512.03385},
      archivePrefix={arXiv},
      primaryClass={cs.CV},
      url={https://arxiv.org/abs/1512.03385}, 
}

@misc{schmitt2025,
      title={Simulating dynamics of correlated matter with neural quantum states}, 
      author={Markus Schmitt and Markus Heyl},
      year={2025},
      eprint={2506.03124},
      archivePrefix={arXiv},
      primaryClass={quant-ph},
      url={https://arxiv.org/abs/2506.03124}, 
}

@article{chen2024,
  title={Empowering deep neural quantum states through efficient optimization},
  author={Chen, Ao and Heyl, Markus},
  journal={Nature Physics},
  volume={20},
  number={9},
  pages={1476--1481},
  year={2024},
  publisher={Nature Publishing Group UK London}
}

@misc{kaltenmark2025,
      title={Spectroscopic signatures of emergent elementary excitations in a kinetically constrained long-range interacting two-dimensional spin system}, 
      author={Tobias Kaltenmark and Chris Nill and Christian Groß and Igor Lesanovsky},
      year={2025},
      eprint={2511.13279},
      archivePrefix={arXiv},
      primaryClass={cond-mat.quant-gas},
      url={https://arxiv.org/abs/2511.13279}, 
}

@Article{Weinberg2017,
	title={{QuSpin: a Python package for dynamics and exact diagonalisation of quantum many body systems part I: spin chains}},
	author={Phillip Weinberg and Marin Bukov},
	journal={SciPost Phys.},
	volume={2},
	pages={003},
	year={2017},
	publisher={SciPost},
	doi={10.21468/SciPostPhys.2.1.003},
	url={https://scipost.org/10.21468/SciPostPhys.2.1.003},
}

@Article{Schmitt2022,
	title={{jVMC: Versatile and performant variational Monte Carlo leveraging automated differentiation and GPU acceleration}},
	author={Markus Schmitt and Moritz Reh},
	journal={SciPost Phys. Codebases},
	pages={2},
	year={2022},
	publisher={SciPost},
	doi={10.21468/SciPostPhysCodeb.2},
	url={https://scipost.org/10.21468/SciPostPhysCodeb.2},
}

@article{yang1989,
  title={Simple variational proof that any two-dimensional potential well supports at least one bound state},
  author={Yang, K and de Llano, M},
  journal={American Journal of Physics},
  volume={57},
  number={1},
  pages={85--86},
  year={1989},
  publisher={American Association of Physics Teachers}
}

@article{chadan2003,
  title={Bound states in one and two spatial dimensions},
  author={Chadan, K and Khuri, NN and Martin, A and Tsun Wu, Tai},
  journal={Journal of Mathematical Physics},
  volume={44},
 number={2},
  pages={406--422},
  year={2003},
  publisher={American Institute of Physics}
}

@article{Yoshinaga2022,
  title = {Emergence of Hilbert Space Fragmentation in Ising Models with a Weak Transverse Field},
  author = {Yoshinaga, Atsuki and Hakoshima, Hideaki and Imoto, Takashi and Matsuzaki, Yuichiro and Hamazaki, Ryusuke},
  journal = {Phys. Rev. Lett.},
  volume = {129},
  issue = {9},
  pages = {090602},
  numpages = {8},
  year = {2022},
  month = {Aug},
  publisher = {American Physical Society},
  doi = {10.1103/PhysRevLett.129.090602},
  url = {https://link.aps.org/doi/10.1103/PhysRevLett.129.090602}
}

@article{Balducci2023,
  title = {Interface dynamics in the two-dimensional quantum Ising model},
  author = {Balducci, Federico and Gambassi, Andrea and Lerose, Alessio and Scardicchio, Antonello and Vanoni, Carlo},
  journal = {Phys. Rev. B},
  volume = {107},
  issue = {2},
  pages = {024306},
  numpages = {28},
  year = {2023},
  month = {Jan},
  publisher = {American Physical Society},
  doi = {10.1103/PhysRevB.107.024306},
  url = {https://link.aps.org/doi/10.1103/PhysRevB.107.024306}
}

@article{Balducci2022,
  title = {Localization and Melting of Interfaces in the Two-Dimensional Quantum Ising Model},
  author = {Balducci, Federico and Gambassi, Andrea and Lerose, Alessio and Scardicchio, Antonello and Vanoni, Carlo},
  journal = {Phys. Rev. Lett.},
  volume = {129},
  issue = {12},
  pages = {120601},
  numpages = {7},
  year = {2022},
  month = {Sep},
  publisher = {American Physical Society},
  doi = {10.1103/PhysRevLett.129.120601},
  url = {https://link.aps.org/doi/10.1103/PhysRevLett.129.120601}
}

@misc{dataset_long-range_quantum_Ising_model,
  author       = {Naik, Vighnesh Dattatraya and
                  Heyl, Markus},
  title        = {Slow dynamics and magnon bound states in the 2D
                   long-range quantum Ising model
                  },
  month        = dec,
  year         = 2025,
  publisher    = {Zenodo},
  doi          = {10.5281/zenodo.17871344},
  url          = {https://doi.org/10.5281/zenodo.17871344},
}
  
\appendix

\onecolumngrid
\vspace{2\baselineskip}  
\begin{center}
\textbf{\large End Matter}
\end{center}
\setcounter{section}{0}
\renewcommand{\thesection}{\Alph{section}}
\vspace{0.5\baselineskip}
\twocolumngrid

\setcounter{equation}{0}
\renewcommand{\theequation}{A\arabic{equation}}

\textit{Appendix A: Neural quantum state simulations.—}We represent the many-body wavefunction by a neural quantum state (NQS)
$\psi_{\boldsymbol{\theta}}(S,t)$, where $S\!\in\!\{\pm1\}^{L\times L}$ and
$\boldsymbol{\theta}$ are trainable parameters.
Real-time dynamics follows the time-dependent variational principle (TDVP),
which projects the Schr\"odinger equation onto the NQS manifold:
\begin{align}
  \sum_{b} S_{ab}\,\dot{\theta}_b \;&=\; -i\,F_a,\\
  \qquad
  S_{ab}=\langle O_a^\dagger O_b\rangle_c,
  &\quad
  F_a=\langle O_a^\dagger E_{\rm loc}\rangle_c ,
  \label{eq:TDVP-A}
\end{align}
with logarithmic derivatives
$O_a(S,t)=\partial_{\theta_a}\ln\psi_{\boldsymbol{\theta}}(S,t)$,
connected correlators $\langle\cdot\rangle_c$ over $|\psi_{\boldsymbol{\theta}}|^2$,
and local energy $E_{\rm loc}(S)=\sum_{S'} H_{SS'}\,\psi_{\boldsymbol{\theta}}(S')/\psi_{\boldsymbol{\theta}}(S)$.
We integrate the equation of motion $S\,\dot{\boldsymbol{\theta}}= -i\mathbf{F}$ using a Heun solver, which provides a second-order update for $\boldsymbol{\theta}(t)$.

We employ a convolutional neural network (CNN) based on the ResNet framework, with Gaussian Error Linear Unit (GELU) activation functions.
We enforce lattice translation invariance by using the convolutions with periodic padding.
The convolutional blocks produce a final set of real-valued feature maps. 
These maps are then split to generate the real and imaginary parts of an intermediate complex feature map.
A global logsumexp pooling operation is applied across all spatial positions and feature channels of this map to produce a base scalar log-wavefunction.
To produce the final NQS, we additionally impose the point-group symmetries of the square lattice (rotations and reflections), by averaging the exponentiated base wavefunction over all symmetric operations at the output layer.
We use networks consisting of two ResNet blocks with $3\times3$ filter kernels and 16 feature channels for all simulations presented in this work.

All simulations are performed in the $\hat{S}^z$ basis, where each configuration $S$ in $\psi_{\boldsymbol{\theta}}(S)$ denotes a spin configuration along the $z$-direction.
Although the Hamiltonian~\ref{eq:Hamiltonian} is written in the $ZZ + X$ form in the main text, we implement the NQS simulations using the equivalent rotated $XX+Z$ representation.
This choice ensures that the ferromagnetically ordered initial state corresponds to a uniformly distributed wavefunction in the configuration space, rather than a single delta peak, which removes instabilities coming from zeroes of the wavefunction.

Expectation values in Eq.~\eqref{eq:TDVP-A} are estimated using Markov-chain Monte Carlo sampling with Metropolis single-spin-flip updates.
For each time step, we use $5 \times L^2$ sweeps for thermal equilibration, followed by $3 \times 10^4 - 4 \times 10^4$ measurement sweeps.
We use an adaptive Heun scheme for the integration of the TDVP equation, with integration tolerance of $10^{-4}$.\\

\textit{Appendix B: Spectral analysis.—} 
To extract excitation energies, we compute the discrete Fourier transform (FFT) of 
$\mathcal O(t)\in\{C(d,t),\,\langle\hat{S}_i^z(t)\rangle\}$ over a finite time and apply a Hamming window to reduce spectral leakage.
With $t_n = n\,\Delta t$ and $n = 0,\ldots,N_t\!-\!1$, we define 
$w_n = 0.54 - 0.46\cos\!\left(\tfrac{2\pi n}{N_t-1}\right)$,
and the transform
\begin{equation}
\mathcal{F}({\mathcal O}(t))(\omega_k)
    = \sum_{n=0}^{N_t-1} 
      w_n\,\big[\mathcal O(t_n)-\overline{\mathcal O}\big]\,
      e^{i\omega_k t_n},
\label{eq:fft}
\end{equation}
where $\overline{\mathcal O}=\tfrac{1}{N_t}\sum_{n=0}^{N_t-1}\mathcal O(t_n)$ removes the zero-frequency component and 
$\omega_k=\tfrac{2\pi k}{N_t\,\Delta t}$ with $k=0,\ldots,N_t\!-\!1$.
Peaks in $|\mathcal{F}({\mathcal O}(t))(\omega_k)|$ identify the frequencies governing the dynamics.
These peak positions are compared to energy differences $|E_i^\nu - E_j^{\nu'}|$ obtained by diagonalizing the effective few-magnon Hamiltonians $\hat H_{\mathrm{eff}}^{(\nu)}$, allowing us to assign each spectral line to a single-magnon or a multi-magnon bound state.
\newpage

\pagebreak

\onecolumngrid

\widetext
\begin{center}
\textbf{\large Supplemental Material: Persistent coherent quantum dynamics in 2D long-range magnets via magnon binding}
\end{center}

\setcounter{equation}{0}
\setcounter{figure}{0}
\setcounter{table}{0}
\setcounter{page}{1}
\makeatletter
\renewcommand{\theequation}{S\arabic{equation}}
\renewcommand{\thefigure}{S\arabic{figure}}
\renewcommand{\bibnumfmt}[1]{[S#1]}
\renewcommand{\citenumfont}[1]{S#1}

\section{Schrieffer-Wolff transformation.}
This section derives the effective few-magnon Hamiltonian using the Schrieffer–Wolff (SW) transformation.
A configuration with $\nu$ magnons at positions ${i_1,\ldots,i_\nu}$ has unperturbed energy
\begin{equation*}
    E_{\nu}(\{i_m\}) = E_{0} + 2\nu J\left(1 - \frac{1}{L^2}\right) - \frac{2J}{\mathcal{N}_\alpha}\sum_{1\le k < l\le\nu} \frac{1}{r^\alpha_{i_ki_l}}.
\end{equation*}
where $E_0 = -(L^2 - 1)/2$ denotes the energy of the fully polarized state and only energy differences enter the SW expansion.
With zero magnon states being the lowest energy band, $E_{\nu}(\{i_m\})$ helps to determine the structure of low-energy excitations.
In the nearest-neighbor limit ($\alpha=\infty$), the energy levels are classified solely by the interface length.
For finite $\alpha$, zero-, one-, and two-magnon sectors form well-separated excitation bands with gaps of order $\mathcal{O}(J)$.  

Introducing a weak transverse field $g\hat T = -\sum_i \hat{S}_i^x$, couples different $\nu$ sectors only perturbatively.
To eliminate these couplings at leading order, we perform the Schrieffer--Wolff (SW) transformation $
\hat H_{\rm SW} = e^{\hat S}\hat H e^{-\hat S}$,
and choose the generator $\hat S$ such that $[\hat S, \hat H_0] = -g\,\hat T,$
which removes matrix elements between the $\nu$ and $\nu\!\pm\!1$ sectors to first order in~$g$.
Projecting onto the $\nu$-magnon manifold $\mathcal P_\nu$ gives
\begin{equation*}
\hat H_{\rm eff}^{(\nu)}
=
P_\nu \hat H_0 P_\nu
+
\frac{g^2}{2}\,
P_\nu [\hat S, \hat T] P_\nu
+
\mathcal O(g^3/J^2).
\end{equation*}
Using the standard expression for $\hat S$ in terms of virtual transitions, we obtain
\begin{equation*}
\hat H_{\rm eff}^{(\nu)} = 
\sum_m E_\nu(m)\, |m\rangle\!\langle m|
\nonumber +
\frac{g^2}{2}
\sum_{m,n,\beta}
\langle m|\hat T|\beta\rangle\,
\langle\beta|\hat T|n\rangle
\left(
\frac{1}{E_m - E_\beta}
+
\frac{1}{E_n - E_\beta}
\right)
|m\rangle\!\langle n|,
\end{equation*}
where $\{|m\rangle\}$ spans the $\nu$-magnon space $\mathcal P_\nu$, and $|\beta\rangle$ runs over virtual states in the complement $\mathcal Q_\nu = 1 - \mathcal P_\nu$ (i.e., the $\nu-1$ and $\nu+1$ sectors).  
This equation provides the leading $\mathcal O(g^2/J)$ corrections that define the effective few-magnon Hamiltonian used in the main text.

\section{Effective zero--magnon Hamiltonian}

In the zero--magnon sector ($\nu = 0$) there is only a single basis state.  
The effective Hamiltonian $\hat H_{\rm eff}^{(0)}$ therefore reduces to the diagonal energy
\begin{equation*}
E_{\rm eff}^{(0)}
=
E_0
+
g^{2}
\sum_{\beta}
\frac{|\langle 0|\hat T|\beta\rangle|^{2}}{E_0 - E_\beta},
\end{equation*}
where $|\beta\rangle$ runs over all single--magnon states, and all single--magnon states lie at energy
$E_\beta = E_0 + 2J\!\left(1 - \frac{1}{L^{2}}\right).$
Summing over the $L^{2}$ possible single--flip configurations gives
\begin{equation*}
\boxed{
E_{\rm eff}^{(0)}
=
E_0 - \frac{g^{2}}{8J}\frac{L^{2}}{(1-1/L^2)} \approx E_0 - \frac{g^{2}L^{2}}{8J}.}
\end{equation*}

\section{Effective single-magnon Hamiltonian.}
In the single–magnon sector ($\nu = 1$), the effective Hamiltonian $\hat H_{\rm eff}^{(1)}$ can be written as
\begin{align*}
\hat H_{\rm eff}^{(1)}
=&\; E_1 \sum_{i} \ket{i}\!\bra{i}
+ g^2
\sum_{i,j,\beta}
\frac{\bra{i}\hat T\ket{\beta}\,
\bra{\beta}\hat T\ket{j}}{E_1-E_\beta}
\ket{i}\!\bra{j},
\end{align*}
where $\ket{i}$ are single–flip states with $i$ denoting the position of the magnon, and
$E_1 = E_0 + 2J\left(1 - \frac{1}{L^2}\right)$.
Here $\ket{\beta}$ is either the zero–magnon state or any of the two–magnon states.
\begin{figure}
    \centering
    \includegraphics[width=\columnwidth]{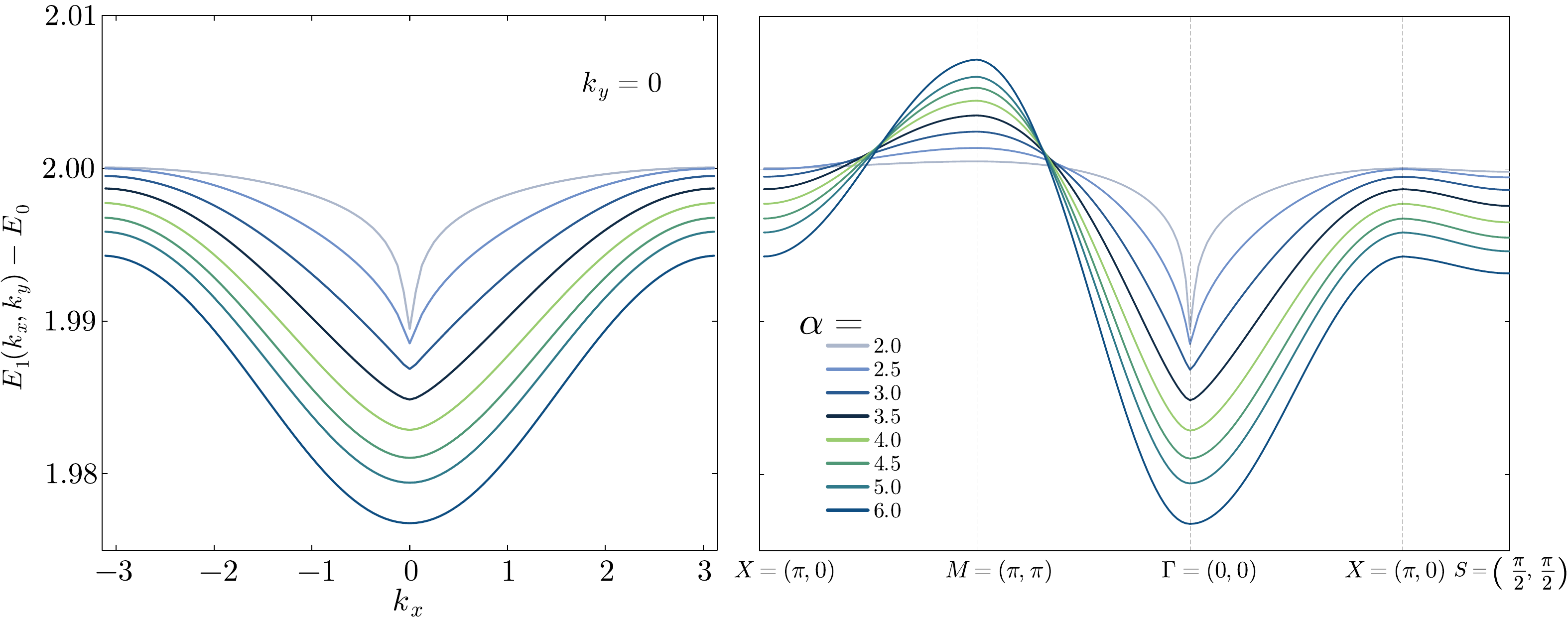}%
    \caption{Single-magnon dispersion for different power-law exponents $\alpha$.
    Left: dispersion $E_1(k_x,k_y{=}0)-E_0$ along the $k_x$ direction.
    Right: dispersion along the high-symmetry path $X\to M\to\Gamma\to X\to S$.
    Long-range interactions induce a nonanalytic cusp at $\Gamma$, which gradually disappears as $\alpha$ becomes large.}
    \label{fig:S1}
\end{figure}

\paragraph{Diagonal term.}
The diagonal term also acquires a second–order correction in $g$ to $E_1$,
\begin{align*}
    U_i
    &= E_0 + 2J\left(1 - \frac{1}{L^2}\right)
    + g^2\sum_{\beta\in (0,2)\text{–flip}} 
    \frac{\bigl|\bra{i}\hat T\ket{\beta}\bigr|^2}{E_1-E_\beta}\\[2pt]
    &= E_0 + 2J\left(1 - \frac{1}{L^2}\right)
    + \frac{g^2}{8J\left(1 - \frac{1}{L^2}\right)}
    - \frac{g^2}{4}\sum_{j\neq i}
    \frac{1}{2J\left(1 - \frac{1}{L^2}\right)
    -\frac{2J}{\mathcal{N}_\alpha\,|r_{ij}|^\alpha}}.
\end{align*}
Taking the large–$L$ limit, and writing $\mathbf{d} = r_{ij}$, this simplifies to
\begin{align*}
    U
    \approx E_0 + 2J
    + \frac{g^2}{8J}
    - \frac{g^2}{8J}\sum_{|\mathbf{d}|\neq 0}
    \left(1 - \frac{1}{\mathcal{N}_\alpha}\frac{1}{|\mathbf{d}|^\alpha}\right)^{-1}.
\end{align*}
Expanding
\(
\left(1 - \frac{1}{\mathcal{N}_\alpha}\frac{1}{|\mathbf{d}|^\alpha}\right)^{-1}
\)
to first order in $\frac{1}{|\mathbf{d}|^\alpha}$, we obtain
\begin{align*}
    U
    &\approx E_0 + 2J
    + \frac{g^2}{8J}
    - \frac{g^2}{8J}\sum_{|\mathbf{d}|\neq 0}
    \left(1 + \frac{1}{\mathcal{N}_\alpha}\frac{1}{|\mathbf{d}|^\alpha}\right)\\[2pt]
    &= E_0 + 2J
    + \frac{g^2}{8J}
    - \frac{g^2(L^2-1)}{8J}
    - \frac{g^2}{8J\mathcal{N}_\alpha}\sum_{|\mathbf{d}|\neq 0}\frac{1}{|\mathbf{d}|^\alpha}\\[2pt]
    &= E_0 + 2J - \frac{g^2L^2}{8J}.
\end{align*}
Therefore, in the large–$L$ limit the diagonal part of $\hat H_{\rm eff}^{(1)}$ is well approximated by
\begin{equation*}
    U \approx E_0 + 2J - \frac{g^2 L^2}{8J}.
\end{equation*}

\paragraph{Off–diagonal term.}
The off–diagonal matrix elements between two single–magnon states are given by
\begin{align*}
t_{i,j}
&= g^2
\sum_{\beta\in (0,2)\text{–flip}}
\frac{
\bra{i}\hat T\ket{\beta}\,\bra{\beta}\hat T\ket{j}
}{E_1-E_\beta}\\[2pt]
&= \frac{g^2}{4}\left(
\frac{1}{2J\left(1 - \frac{1}{L^2}\right)}    
-
\frac{1}{2J\left(1 - \frac{1}{L^2}\right)
-\frac{2J}{\,|r_{ij}|^\alpha}}
\right).
\end{align*}
Taking again the large–$L$ limit and expanding to first order in $\frac{1}{|r_{ij}|^\alpha}$, we obtain
\begin{align*}
    t_{i,j} \approx -\frac{g^2}{8J\mathcal{N}_\alpha}\left(\frac{1}{|r_{ij}|^\alpha-1/\mathcal{N}_\alpha}\right)
    \approx -\frac{g^2}{8J\mathcal{N}_\alpha}\,\frac{1}{|r_{ij}|^\alpha}.
\end{align*}
Combining the diagonal and off–diagonal processes, the effective single–magnon Hamiltonian becomes
\begin{align*}
\boxed{
\hat H_{\rm eff}^{(1)} =
\left(E_0 + 2J - \frac{g^2L^2}{8J}\right)
\;-\;
\frac{g^2}{8J\mathcal{N}_\alpha}
\sum_{i \neq j}\frac{1}{|r_{ij}|^\alpha}
\ket{i}\!\bra{j}.}
\end{align*}
Figure~\ref{fig:S1} displays the single-magnon dispersion for different power-law exponents.

\section{Effective two-magnon Hamiltonian.}
In the two–magnon sector ($\nu = 2$), the effective Hamiltonian $\hat H_{\rm eff}^{(2)}$ can be written as
\begin{align*}
\hat H_{\rm eff}^{(2)}
=&\;
\sum_{n_1,n_2} E_2(n_1,n_2)\,\ket{n_1,n_2}\!\bra{n_1,n_2} +
\frac{g^2}{2}
\sum_{n_1,n_2,m_1,m_2,\beta}
\bra{n_1,n_2}\hat T\ket{\beta}\,
\bra{\beta}\hat T\ket{m_1,m_2}\\
&\quad\quad\quad\quad\quad\quad\quad\times
\Bigl(
\frac{1}{E_2(n_1,n_2)-E_\beta}+
\frac{1}{E_2(m_1,m_2)-E_\beta}
\Bigr)
\ket{n_1,n_2}\!\bra{m_1,m_2}.
\end{align*}
Here $\ket{n_1,n_2}$ are two–flip states inside the two–magnon subspace $\mathcal P$, while $\ket{\beta}$ are virtual states in the complement $\mathcal Q = 1-\mathcal P$ (i.e., the one–flip and three–flip sectors), and
\[
E_2(n_1,n_2)
=
E_0
+
4J\!\left(1-\frac{1}{L^2}\right)
-
\frac{2J}{\mathcal{N}_\alpha}\frac{1}{r^\alpha_{n_1,n_2}}.
\]

\paragraph{Diagonal term.}
We now switch to relative coordinates by pinning one flipped spin and labeling basis states by the displacement
$\mathbf d = \mathbf n_2 - \mathbf n_1$, so that $\ket{\mathbf d}$ denotes a two–magnon state with separation $\mathbf d$.
The diagonal term then acquires a second–order correction in $g$,
\begin{align*}
    U(\mathbf d)
&= E_0 + 4J\!\left(1-\frac{1}{L^2}\right)
   - \frac{2J}{\mathcal{N}_\alpha}\frac{1}{|\mathbf{d}|^\alpha}
   +
   g^2
   \sum_{\beta\in (1,3)\text{–flip}}
   \frac{|\langle \mathbf d|\hat T|\beta\rangle|^2}{E_2(\mathbf d)-E_\beta} \\
&= E_0 + 4J\!\left(1-\frac{1}{L^2}\right)
   - \frac{2J}{\mathcal{N}_\alpha}\frac{1}{|\mathbf{d}|^\alpha} \\
&\quad
   + \frac{g^2}{4} \left[
   \frac{2}{2J\left(1 - \frac{1}{L^2}\right)
        - \frac{2J}{\mathcal{N}_\alpha}\frac{1}{|\mathbf{d}|^\alpha}}
   - \sum_{\substack{\mathbf{d'} \neq(0,0) \\ \mathbf{d'} \neq \mathbf{d}}}
   \frac{1}{2J\left(1 - \frac{1}{L^2}\right)
   - \frac{2J}{\mathcal{N}_\alpha}\left[\frac{1}{|\mathbf{d'}|^\alpha}
   + \frac{1}{|\mathbf{d} - \mathbf{d'}|^\alpha}\right]}
   \right].
\end{align*}

Taking the large-$L$ limit gives the leading–order corrections.
The first correction term can be written as
\begin{align*}
    \frac{g^2}{8J} \frac{2}{\left(1 - \frac{1}{\mathcal{N}_\alpha}\frac{1}{|\mathbf{d}|^\alpha}\right)}
    &= \frac{g^2}{4J} \left[1 - \frac{1}{\mathcal{N}_\alpha}\frac{1}{|\mathbf{d}|^\alpha}\right]^{-1} 
    \approx \frac{g^2}{4J} \left[1 + \frac{1}{\mathcal{N}_\alpha}\frac{1}{|\mathbf{d}|^\alpha}\right] = \frac{g^2}{4J} + \frac{g^2}{4J\mathcal{N}_\alpha}\frac{1}{|\mathbf{d}|^\alpha}.
\end{align*}

The second correction term is
\begin{align*}
    \frac{g^2}{8J}
    \sum_{\substack{\mathbf{d'} \neq(0,0) \\ \mathbf{d'} \neq \mathbf{d}}}
    \frac{1}{\left[1 - \frac{1}{\mathcal{N}_\alpha}\left(\frac{1}{|\mathbf{d'}|^\alpha}
    + \frac{1}{|\mathbf{d} - \mathbf{d'}|^\alpha}\right)\right]}
    &= \frac{g^2}{8J}
    \sum_{\substack{\mathbf{d'} \neq(0,0) \\ \mathbf{d'} \neq \mathbf{d}}}
    \left[1 + \sum^{\infty}_{i=1}\left(\frac{1}{\mathcal{N}_\alpha}\right)^i
    \left(\frac{1}{|\mathbf{d'}|^\alpha}
    + \frac{1}{|\mathbf{d} - \mathbf{d'}|^\alpha}\right)^i\right] \\
    &= \frac{g^2(L^2 - 2)}{8J}
    + \frac{g^2}{8J}
    \sum^{\infty}_{i=1}\left[\left(\frac{1}{\mathcal{N}_\alpha}\right)^i
    \sum_{\substack{\mathbf{d'} \neq(0,0) \\ \mathbf{d'} \neq \mathbf{d}}}
    \left(\frac{1}{|\mathbf{d'}|^\alpha}+
    \frac{1}{|\mathbf{d} - \mathbf{d'}|^\alpha}\right)^i\right] \\
    &= \frac{g^2(L^2 - 2)}{8J}
    + \frac{g^2}{8J}
    \sum^{\infty}_{i=1}\left[
    \left(\frac{1} {\mathcal{N}_\alpha}\right)^i
    \sum_{m=0}^{i} \binom{i}{m}
    \sum_{\substack{\mathbf{d'} \neq(0,0) \\ \mathbf{d'} \neq \mathbf{d}}} \frac{1}{|\mathbf{d'}|^{m\alpha}|\mathbf{d} - \mathbf{d'}|^{(i-m)\alpha}}\right].
\end{align*}

Separating the $\mathbf{d}$–dependent terms, we obtain
\begin{align*}
    \frac{g^2}{8J}
    \sum_{\substack{\mathbf{d'} \neq(0,0) \\ \mathbf{d'} \neq \mathbf{d}}}
    \frac{1}{\left[1 - \frac{1}{\mathcal{N}_\alpha}
    \left(\frac{1}{|\mathbf{d'}|^\alpha}
    + \frac{1}{|\mathbf{d} - \mathbf{d'}|^\alpha}\right)\right]}
    &= \frac{g^2(L^2-2)}{8J}
    + \frac{g^2}{4J}\sum_{i = 1}^{\infty}\left(\frac{1}{\mathcal{N}_\alpha}\right)^i
      \left(\sum_{\mathbf{d'} \neq (0,0)}
      \frac{1}{|\mathbf{d'}|^{i\alpha}}\right) \\
    &\quad
    - \frac{g^2}{4J}\sum_{i = 1}^{\infty}\left(\frac{1}{\mathcal{N}_\alpha}\right)^i
      \frac{1}{|\mathbf{d}|^{i\alpha}}
    + \frac{g^2}{8J}\sum_{i = 2}^{\infty}\left(\frac{1}{\mathcal{N}_\alpha}\right)^i
      \sum^{i-1}_{m=2}\mathcal{F}^i_m(\mathbf{d}),
\end{align*}
where
\begin{equation*}
    \mathcal{F}^i_m(\mathbf{d})
    =
    \sum_{\substack{\mathbf{d'} \neq (0,0) \\ \mathbf{d'} \neq \mathbf{d}}}
    \frac{1}{|\mathbf{d'}|^{m\alpha}|\mathbf{d} - \mathbf{d'}|^{(i-m)\alpha}}
\end{equation*}
can be evaluated for large $|\mathbf d|$ by approximating the sum by a continuum integral,
\begin{equation*}
    \mathcal{F}^i_m(R)
    =
    \int d^2r\,\frac{1}{|r|^{m\alpha}|r-R|^{(i-m)\alpha}}.
\end{equation*}
Using the homogeneous function property, one finds
\begin{align*}
    \mathcal{F}^i_m(\lambda R)
    = \frac{1}{\lambda^{(i\alpha - 2)}}\mathcal{F}^i_m(R)
    \quad\Rightarrow\quad
    \mathcal{F}^i_m(R)
    = \frac{\mathcal{F}^i_m(1)}{R^{(i\alpha - 2)}}.
\end{align*}

Having all these ingredients, we can now write down the second correction term, ignoring higher–order contributions in $i$:
\begin{align*}
    \frac{g^2(L^2+2)}{8J}
    - \frac{g^2}{4J\mathcal{N}_\alpha}\frac{1}{|\mathbf{d}|^\alpha}
    + \frac{g^2}{8J\mathcal{N}^2_\alpha}
      \frac{\mathcal{F}^{i=2}_{m=1}(1)}{|\mathbf{d}|^{2(\alpha - 1)}}.
\end{align*}

Finally, combining everything gives the expression for $U(\mathbf{d})$:
\begin{equation*}
    \tilde{U}(\mathbf{d})
    =
    E_0 + 4J
    - \frac{g^2L^2}{8J}
    - \frac{(2J - g^2/2J)}{\mathcal{N}_\alpha}\frac{1}{|\mathbf{d}|^\alpha},
\end{equation*}
and we isolate the $\mathbf d$–dependent potential as
\begin{equation*}
    U(\mathbf{d})
    =
    - \frac{(2J - g^2/2J)}{\mathcal{N}_\alpha}\frac{1}{|\mathbf{d}|^\alpha}.
\end{equation*}

\paragraph{Off–diagonal term.}
The off–diagonal matrix elements between distinct separations are
\begin{align*}
t_{\mathbf d, \mathbf d'}
&=
\frac{g^2}{2}
\sum_{\beta\in (1,3)\text{–flip}}
\bra{\mathbf d}\hat T\ket{\beta}\,
\bra{\beta}\hat T\ket{\mathbf d'}
\Bigl(
\frac{1}{E_2(\mathbf d)-E_\beta}
+
\frac{1}{E_2(\mathbf d')-E_\beta}
\Bigr)\\
&=
\frac{g^2}{8} \left(
\frac{1}{2J(1-1/L^2) - \frac{2J}{\mathcal{N}_\alpha}\frac{1}{|\mathbf{d}|^\alpha}}
+ \frac{1}{2J(1-1/L^2) - \frac{2J}{\mathcal{N}_\alpha}\frac{1}{|\mathbf{d'}|^\alpha}}
\right)\\
&\quad
- \frac{g^2}{8}\left(
\frac{1}{2J(1-1/L^2)
- \frac{2J}{\mathcal{N}_\alpha}\left(\frac{1}{|\mathbf{d'}|^\alpha}
+ \frac{1}{|\mathbf{d-d'}|^\alpha}\right)}
+ \frac{1}{2J(1-1/L^2)
- \frac{2J}{\mathcal{N}_\alpha}\left(\frac{1}{|\mathbf{d}|^\alpha}
+ \frac{1}{|\mathbf{d-d'}|^\alpha}\right)}
\right).
\end{align*}
Taking the large-$L$ limit and expanding to first order in $\frac{1}{|\mathbf{d}|^\alpha}$, we get
\begin{align*}
t_{\mathbf d, \mathbf d'}
&\approx \frac{g^2}{16J}\Biggl(
1 + \frac{1}{\mathcal{N}_\alpha}\frac{1}{|\mathbf{d}|^\alpha}
+ 1 + \frac{1}{\mathcal{N}_\alpha}\frac{1}{|\mathbf{d'}|^\alpha} \\
&\qquad\qquad
- 1 - \frac{1}{\mathcal{N}_\alpha}\left(\frac{1}{|\mathbf{d}|^\alpha}
+ \frac{1}{|\mathbf{d}-\mathbf{d'}|^\alpha}\right)
- 1 - \frac{1}{\mathcal{N}_\alpha}\left(\frac{1}{|\mathbf{d'}|^\alpha}
+ \frac{1}{|\mathbf{d}-\mathbf{d'}|^\alpha}\right)
\Biggr),
\end{align*}
which simplifies to
\begin{align*}
t_{\mathbf d, \mathbf d'}
\approx -\frac{g^2}{8J\mathcal{N}_\alpha}\frac{1}{|\mathbf{d}-\mathbf{d'}|^\alpha}.
\end{align*}

Combining the diagonal and off–diagonal processes gives
\begin{align*}
\boxed{
\hat H^{(2)}_{\mathrm{eff}}
=
E_2
+
\sum_{\mathbf d}U(\mathbf d)\,\ket{\mathbf d}\!\bra{\mathbf d}
+
\sum_{\mathbf d \neq \mathbf d'}
t_{\mathbf d, \mathbf d'}\,\ket{\mathbf d}\!\bra{\mathbf d'},}
\end{align*}
with 
\begin{align*}
    E_2 &\approx E_0 + 4J - \frac{g^2L^2}{8J}, \qquad
    U(\mathbf{d}) \approx -\frac{(2J - g^2/2J)}{\mathcal{N}_\alpha}\frac{1}{|\mathbf{d}|^\alpha}, \qquad
    t_{\mathbf d, \mathbf d'} \approx -\frac{g^2}{8J\mathcal{N}_\alpha}\frac{1}{|\mathbf{d}-\mathbf{d'}|^\alpha}.
\end{align*}

\end{document}